\documentstyle[preprint,aps]{revtex}

\begin{document}
\bibliographystyle{prsty}
\draft

\title{Series solution to the laser-ion interaction in a Raman-type
configuration}
\author{Mang Feng
\thanks{Present address: Institute for Scientific Interchange Foundation, Villa Gualino, Viale Settimio 
Severo 65, I-10131, Torino, Italy. Electronic address: feng@isiosf.isi.it}}
\address{$^{1}$Max-Planck Institute for the Physics of Complex Systems,\\
N$\ddot{o}$thnitzer Street 38, 01187 Dresden, Germany\\
$^{2}$Laboratory of Magnetic Resonance and Atomic and Molecular Physics,\\ 
Wuhan Institute of Physics and Mathematics, Academia Sinica,\\
Wuhan, 430071, People's Republic of China}

\date{\today}
\maketitle

\begin{abstract}

The Raman interaction of a trapped ultracold ion with two travelling wave 
lasers is studied analytically with series solutions, in the absence of the 
rotating wave approximation (RWA) and the restriction of both the Lamb-Dicke limit 
and the weak excitation regime. The comparison is made between our solutions and those 
under the RWA to demonstrate the validity region of the RWA. As a practical example, the preparation of 
Schr\"odinger-cat states with our solutions is proposed beyond the weak excitation regime.
\end{abstract}
\vskip 1cm

\narrowtext
\section{introduction}
The preparation of ultracold ions as well as the
production of nonclassical motional states of these ions plays a central role in ion trap 
experiments$^{[1-3]}$. As the trapped ion
system is in  nearly perfect isolation, immune from the violation of the
environment, and meanwhile the electric-dipole forbidden transition is usually
introduced for avoiding the dissipation of the ion, the decoherence in the
preparation and preservation of nonclassical states can be
effectively suppressed. Consequently, the ion trap system is also being considered 
to be a promising tool for the quantum computing$^{[4-6]}$.

The various theoretical schemes for generating the nonclassical states of motion of the 
trapped ions and achieving quantum computing with trapped ions are based on the
Jaynes-Cummings (JC) model$^{[7]}$, in which the ion is simplified to be of two
levels, the trap's potential is quantized as a harmonic oscillator, and
the radiating lasers are supposed to be classical forms of  standing or travelling waves.
The general consideration is taken for the case of Lamb-Dicke limit (LDL) under the weak excitation regime (WER),
which corresponds to the actual case in present ion trap experiments$^{[3]}$.
The LDL means the ions moving within the region of the space much smaller than the effective wavelength of the 
laser wave, and the WER means the laser-ion interaction much smaller than the trap frequency. For this case, 
some techniques developed in the framework of cavity QED based on JC model can be immediately transcribed 
to the ion trap system by taking advantage of the analogy between the cavity QED and  ion trap problem.
In the case of motion of ions exceeding the LDL, we can use the technique proposed in Ref.[8] for a 
treatment. In a word, under the rotating-wave approximation (RWA), the problem of trapped ions 
interacting with lasers is solvable.  
However, the strength of the coupling between the ions and lasers can be conveniently adjusted simply 
by changing the intensity of lasers.  If  the coupling strength is much larger than the trap frequency, 
called the strong excitation regime (SER), then the RWA is no longer applicable since the rapidly 
oscillating (i.e., counter-rotating) terms also make significant contribution in the interaction, which 
is much different from the case in Cavity QED$^{[9]}$. It has been reported
that  the SER is useful for preparing nonclassical motional states of a trapped ion and implementing quantum 
computing with trapped ions$^{[6,10-12]}$. For example in Ref.[10], Schr\"odinger cat states 
could be prepared readily in simple way in the SER. Based on that work, a proposal was made for motional state 
engineering and endoscopy  in the SER$^{[11]}$. As the quantum state is sensitive to  decoherence, the 
manipulation of trapped ions in the SER is of importance to the
experimental implementation and measurement because of the great reduction of the operation time. Due to the same 
reason, a scheme for the hot-ion quantum computing also requires the condition of the SER$^{[6]}$. 
In contrast to the theoretical studies, the ion-laser interaction in the SER has not yet been reported 
experimentally
since the high-laser intensity produces increased off-resonant excitation of carrier transitions which
limits the final ground state occupation of the cooling process$^{[13]}$, and the strong laser also modifies the original 
electronic levels of the ions$^{[14]}$, making the problem somewhat complicated. In a numerical study of the strong 
Raman sideband excitation driving the vibrational ground state of the trapped ion$^{[15]}$, it is shown that the 
non-RWA interaction terms produce some limitations for two strong laser beams detuned by the vibrational
frequency to be used as Raman motional displacement beams, whereas the interaction is  useful for the preparation of 
the Schr\"odinger cat state.

Generally speaking, as the RWA is invalid,  the ion trap problem in the SER can not be solved analytically with JC model. So some 
approximations have to be introduced in Refs.[10,11] and the studies in Refs.[12,15] are resorted to numerical calculations. 
In fact, as shown in Ref.[16], even if the ions are governed in the WER, in which the RWA is valid, the deletion 
of the RWA can also present some interesting results much different from those under the RWA.  In that work, 
some particular analytical solutions have been obtained in the absence of the RWA for large detuning cases, 
and Schr\"odinger cat states  
could be readily prepared. However, that work is too simple. We hope to investigate the Raman configuration which has been extensively 
applied in actual ion trap experiments$^{[3]}$, instead of the simple case in Ref.[16]. We also hope to exclude the 
assumption related to both the WER and LDL. Therefore, in this contribution,  we will try to undertake a more general and complete 
investigation on the trapped ion under the Raman process. Moreover, a comparison of our results with those under the RWA will be made.
Besides, we will propose a scheme for preparing the Schr\"odinger cat state beyond the WER.

\section{model and solution}

We study the single ultracold ion radiated by lasers in Raman-$\Lambda$-type configuration, as shown in Fig.1. 
The electronic structure is employed with two lower levels $|e>$ and $|g>$ coupled to a common upper 
state $|r>$, and the two lasers with frequencies $\omega_{1}$ and 
$\omega_{2}$ respectively are assumed to propagate along opposite directions. For a 
sufficiently large detuning to the level $|r>$, $|r>$ may be adiabatically 
eliminated, and what we have to treat is an effective two-level system, in which the 
lasers drive the
electric-dipole forbidden transition $|g>$$\leftrightarrow |e>$.
The dimensionless Hamiltonian of such a system in the frame rotating with the 
effective laser frequency $\omega_{l}$$(=\omega_{1}-\omega_{2})$ can be
written as$^{[10]}$
\begin{equation}
H=\frac {\Delta}{2} \sigma_{z} + a^{+}a
+\frac {\Omega}{2}(\sigma_{+}e^{i\eta \hat{x}}+\sigma_{-}e^{-i\eta \hat{x}})
\end{equation}
where the detuning $\Delta=(\omega_{0}-\omega_{l})/\nu$ with $\omega_{0}$ 
being the optical resonance frequency, i.e., the transition frequency of two 
levels of the ion, and $\nu$ the frequency of the trap. $\Omega$ is the dimensionless Rabi 
frequency and $\eta$
the effective Lamb-Dicke parameter given by $\eta=\eta_{1}+\eta_{2}$ with 
subscripts denoting the counterpropagating laser fields. $\sigma_{i}$ ($i=\pm, z$) are Pauli operators,
$\hat{x}=a^{+}+a$ is the dimensionless position operator of the ion with
$a^{+}$ and $a$ being operators of creation and annihilation of the phonon 
field, respectively. The notations '+' and '-' in front of $i\eta \hat{x}$ 
indicate the absorption of a photon from one beam followed by emission into the other beam and vice versa,
respectively. $\nu$ is generally supposed to be much greater 
than the atomic decay rate, called the strong confinement limit, for neglecting the effect 
of the atomic decay. In general, one may expand $e^{\pm i\eta\hat{x}}$
in Eq.(1) to the first-order terms of $\eta\hat{x}$ and neglect other
higher-order terms by supposing the ion governed under the 
WER ($\Omega\ll 1$) and within the LDL ($\eta\ll 1$).
However, as we hope to treat the problem more exactly and generally,
we first perform following unitary transformations on Eq.(1), that is
\begin{equation}
H^{I}=V^{+}U^{+}HUV=\frac {\Omega}{2}\sigma_{z} + a^{+}a+ g(a^{+}+a)(\sigma_{+}+
\sigma_{-})+\epsilon (\sigma_{+}+\sigma_{-})+g^{2}
\end{equation}
where $U=\pmatrix{D(\beta) & -D(\beta)\cr D^{+}(\beta)& D^{+}(\beta)}$ and 
$ V=\frac {1}{\sqrt{2}}e^{-i\pi a^{+}a/2}$ with $D(\beta)=e^{i\eta(a^{+}+a)/2}$, $g=\eta/2$ and 
$\epsilon=-\Delta/2$. Actually, the unitary transformation $U$ made above is identical to
that in [17]. But for coinciding with the standard form of non-RWA model, $V$ is performed.
Comparing with the standard non-RWA JC model$^{[9]}$, there exist an additional 
driven term and an additional constant term in Eq.(2), and the optical resonance frequency is
replaced by the Rabi frequency. In what follows, we will adopt the coherent 
state representation$^{[18]}$, by which the non-RWA JC models have been
analytically treated$^{[16,19]}$. Thus Eq.(2) is rewritten as
\begin{equation}
H={\frac {\Omega}{2}}\sigma_{z}+  \alpha {\frac {d}{d \alpha}}+
g(\alpha+\frac {d}{d \alpha})(\sigma_{+}+\sigma_{-})
+\epsilon (\sigma_{+}+\sigma_{-})+g^{2}
\end{equation}
where $\alpha$ is a complex number, and $\int \frac {d\alpha d\alpha^{*}}
{2\pi i} \exp (-|\alpha|^{2}) |\alpha^{*}><\alpha^{*}|=1$ with the relation
between the coherent state and Fock state 
\begin{equation}
|n>=\int \frac {d\alpha d\alpha^{*}}{2\pi i} 
\exp (-|\alpha|^{2}) |\alpha^{*}>\frac {1}{\sqrt{n!}} \alpha^{n}.
\end{equation}
Using the same idea as in Refs.[16,19], we assume the eigenfunction of the 
Schr\"odinger equation of Eq.(3) to be the series form
$$\Psi(\alpha)= \pmatrix
{\Psi_{1}(\alpha) \cr \Psi_{2}(\alpha)}
= \pmatrix{exp(-z\alpha)\sum_{n=0}^{\infty} b_{n}\alpha^{n}\cr
exp(-z\alpha)\sum_{n=0}^{\infty} c_{n}\alpha^{n}}$$
where $z$, $b_{n}$ and $c_{n}$ are constants determined later. Taking
$\Psi(\alpha)$ into the Schr\"odinger equation of Eq.(3) will yield two 
recurrence relations
\begin{equation}
b_{n+1}={\frac {1}{g(n+1)}}[(E+{\frac {\Omega}{2}}
-n-g^{2})c_{n}+(gz-\epsilon)b_{n}-gb_{n-1}+ zc_{n-1}],
\end{equation}
\begin{equation}
c_{n+1}={\frac {1}{g(n+1)}}[(E-{\frac {\Omega}{2}}
-n-g^{2})b_{n}+(gz-\epsilon)c_{n}-gc_{n-1}+ zb_{n-1}]
\end{equation}
where $E$ is the trial solution of the eigenenergy of Eq.(3). 
As the procedure of solution is similar to that in Ref.[16], in what follows, we just list the results:

(1) $b_{n}=c_{n}=0$ for $n\geq 2$, we have two solutions. One is $E_{1}=1+\epsilon$ with $E_{1}$ the eigenenergy 
of the system in this case, corresponding to $z= g$ and  a restricted relation of $\Omega=2\sqrt{1+2\epsilon-4g^{2}}$.
The other is $E^{'}_{1}=1-\epsilon$, with $z= -g$ and  a restricted relation of $\Omega=2\sqrt{1-2\epsilon-4g^{2}}$.
The eigenfunctions for above two cases are respectively
$\Psi(\alpha)=e^{-g\alpha} \pmatrix{b_{0}+ b_{1}\alpha\cr
c_{0}+ b_{1}\alpha}$ and $\Psi^{'}(\alpha)=e^{g\alpha} 
\pmatrix{b_{0}^{'}+ b^{'}_{1}\alpha \cr
c^{'}_{0}- b^{'}_{1}\alpha}$, where $c_{0}=\frac {1-\Omega/2+2\epsilon-2g^{2}}
{1+\Omega/2+2\epsilon-2g^{2}}b_{0}$, $b_{1}=\frac {1}{g}
[(1+\frac {\Omega}{2}-g^{2}+\epsilon)c_{0}+(g^{2}
-\epsilon)b_{0}]$, $c^{'}_{0}=-\frac {1-\Omega/2-2\epsilon-2g^{2}}
{1+\Omega/2-2\epsilon-2g^{2}}b_{0}^{'}$, and $b^{'}_{1}=\frac {1}{g}
[(1-\epsilon-g^{2}+\frac {\Omega}{2})c^{'}_{0}-
(g^{2}+\epsilon)b^{'}_{0}]$. $b_{0}$ and $b^{'}_{0}$ can be set to be 1 for
normalisation.

(2) $b_{n}=c_{n}=0$ for $n \geq 3$, the two solutions are $E_{2}=2 \pm\epsilon$ 
corresponding to $z= \pm g$ and two independent restricted conditions, which can be solved from the following 
equation, respectively
$$\frac {2g^{2}-(1-2g^{2}\pm 2\epsilon)(2+\Omega/2-2g^{2}\pm 2\epsilon)
+\Omega+\Omega^{2}/4}
{2g^{2}-(1-2g^{2}\pm 2\epsilon)(2-\Omega/2-2g^{2}\pm 2\epsilon)-\Omega+\Omega^{2}/4}=$$
\begin{equation}
\frac {4g^{2}(2+3\Omega /4-2g^{2}\pm 2\epsilon)-\Omega(g^{2}\mp\epsilon)(3\pm 2\epsilon+\Omega-2g^{2})}
{4g^{2}(2-3\Omega /4-2g^{2}\pm 2\epsilon)+\Omega[(1+\Omega/2)(2-\Omega/2)+ 2(g^{2}\mp \epsilon)^{2}
-3(g^{2}\mp \epsilon)]}
\end{equation}
The eigenfunctions at this moment in the unit of $b_{0}=b^{'}_{0}=1$ are, respectively, 
$$ \Psi(\alpha)=e^{-g\alpha}\pmatrix{1+ b_{1}\alpha + b_{2}\alpha^{2} \cr
c_{0}+ c_{1}\alpha +b_{2}\alpha^{2}}$$
and 
$$\Psi^{'}(\alpha)=e^{g\alpha} 
\pmatrix{1+ b^{'}_{1}\alpha +b^{'}_{2}\alpha^{2}\cr
c^{'}_{0}+ c^{'}_{1}\alpha -b^{'}_{2}\alpha^{2}}$$
where
$c_{0}=\frac {(1-2g^{2}+ 2\epsilon)(2-\Omega/2-2g^{2}+ 2\epsilon)
+\Omega-\Omega^{2}/4-2g^{2}}
{(1-2g^{2}+2\epsilon)(2+\Omega/2-2g^{2}+ 2\epsilon)-\Omega-\Omega^{2}/4
-2g^{2}}$, $c_{1}=\frac {1}{g}[2+\epsilon-g^{2}-\frac {\Omega}{2}+
(g^{2}-\epsilon)c_{0}]$, $b_{1}=\frac {1}{g}[(2+\epsilon-g^{2}
+\frac {\Omega}{2})c_{0}+g^{2}-\epsilon]$, 
$b_{2}=\frac {1}{2g^{2}}\{(g^{2}-\epsilon)(3-\Omega+2\epsilon-2g^{2})+
g^{2}+[(1+\epsilon-g^{2})^{2}-1-\frac {\Omega^{2}}{4}-\frac
{\Omega}{2}-g^{2}-(g^{2}-\epsilon)^{2}]c_{0} \}$,
$c^{'}_{0}=\frac {2g^{2}-(1-2g^{2}- 2\epsilon)(2-\Omega/2-2g^{2}-2\epsilon)
-\Omega+\Omega^{2}/4}
{(1-2g^{2}-2\epsilon)(2+\Omega/2-2g^{2}-2\epsilon)-\Omega-\Omega^{2}/4
-2g^{2}}$, $c^{'}_{1}=\frac {1}{g}[2-\epsilon-g^{2}-\frac {\Omega}{2}-
(g^{2}+\epsilon)c^{'}_{0}]$, $b^{'}_{1}=\frac {1}{g}[(2-\epsilon-g^{2}+
\frac {\Omega}{2})c^{'}_{0}- g^{2}+\epsilon]$, and
$b^{'}_{2}=\frac {1}{2g^{2}}\{(g^{2}+\epsilon)(3-\Omega-
2\epsilon-2g^{2})+g^{2}+[(1-\epsilon-g^{2})^{2}-1-\frac {\Omega^{2}}{4}-\frac
{\Omega}{2}-g^{2}-(g^{2}+\epsilon)^{2}]c^{'}_{0} \}$;

(3) when above technique is extended to the case of 
$b_{n}=c_{n}=0$ for $n \geq N+1$, the series solution becomes very complicated. However the eigenenergy of the system 
is still of the simple form, that is, $E_{N}=N \pm\epsilon$, corresponding to $z= \pm g$ and two independent sets 
of much complicated expressions of restricted conditions which we have to omit here, respectively. 

\section{discussion}

As limitations for both the Rabi frequency and Lamb-Dicke parameter as
well as the RWA are excluded in the present treatment, the results we obtained above 
are more exact and general than those depending on the RWA or the approximate 
expansion of small values of $\eta$ and $\Omega$. However, as the problem 
involving the
counter-rotating terms is non-integrable$^{[16,18,19]}$, the present solutions 
obtained by the termination of the series of the trial solution $\Psi(\alpha)$
are some particular ones, along with some restricted conditions related to 
some parameters of the system.  Nevertheless, with the present solutions,
we can investigate some cases in a wide range of parameters,
particularly for the case beyond the WER and LDL. It is also of interest to make a comparison 
of our results with those under the RWA. Let us first consider a special case with both 
$\Omega \sim \epsilon \ge 1$ and $\eta\rightarrow 0$. In such a case, Eq.(3) is reduced to 
$$H^{s}={\frac {\Omega}{2}}\sigma_{z}+  \alpha {\frac {d}{d \alpha}}
+\epsilon (\sigma_{+}+\sigma_{-}).$$ 
Repeating the above procedure of the series solution for the simplest case, i.e., $b_{n}=c_{n}=0$ for 
$n\geq 2$, we can find $z=0$ and 
$\Psi^{s}=\pmatrix{b_{0}+b_{1}\alpha \cr c_{0}+c_{1}\alpha}$
with $c_{0}=\frac {E^{s}-\Omega/2-1}{\epsilon}b_{0}$, $c_{1}=\frac {E^{s}-\Omega/2-1}{\epsilon}b_{1}$ and 
$E^{s}=1\pm\sqrt{\epsilon^{2}+\frac {\Omega^{2}}{4}}$. We can easily find that the eigenenergy $E^{s}$ is in  
good agreement with the solution in a Fock state representation for $n=1$. However, to make a comparison for 
eigenfunctions with former solutions, we have to return to the 
original representation before Eq.(2). The eigenfunction in the original representation is 
$$\Psi^{os} = UV \Psi^{s} \approx \frac {1}{\sqrt{2}}
\pmatrix{1& -1 \cr 1& 1}\Psi^{s}(\alpha)e^{-i\pi a^{+}a/2}$$
\begin{equation}
=\frac {1}{\sqrt{2}}\pmatrix{(b_{0}-c_{0})|0>-i(b_{1}+c_{1})|1> \cr 
(b_{0}+c_{0})|0>-i(b_{1}+c_{1})|1>}
\end{equation}
where we have used the fact$^{[16,18]}$ that $e^{g\alpha}\alpha^{n}$ in a coherent state representation corresponds 
to the displaced
Fock states $|n,g>$ by means of Eq.(4) and Baker-Campbell-Hausdorff formula. Therefore, in the case of $\Omega \ll 1$ but 
$\epsilon\gg 1$, we have from Eq.(8)
$\Psi^{os}\sim \pmatrix{1 \cr 0}$ or $\pmatrix{0 \cr 1}$, which means no transitions existing in the large detuning case. 
If we assume $\epsilon \rightarrow 0$ and $\Omega$ being arbitrary positive real number, we have 
$\Psi^{os}\sim \pmatrix{1 \cr 1}$, which corresponds to the carrier excitation.
Both solutions above are also in good agreement with the solutions in a Fock state representation$^{[12]}$. However, 
from Eq.(8) we can know that, different from former solutions for the WER case,
the probability amplitudes of up and down states of the ion are different and complicated in the case of the SER 
with large detunings. This is a new result which has not been obtained before by general approaches made in a Fock 
state or dressed state representation.
 
For a more general comparison, we should first present the solutions under the RWA. As referred to in Sec.I, Eq.(1) 
can be treated to be a nonlinear JC model with
the approach proposed in Ref.[8] under the RWA. However, to make the comparison more easy and clear, we start
our RWA treatment from Eq.(2). By performing a unitary
transformation $\exp [-i(\frac {\Omega}{2}\sigma_{z}+\frac {a^{+}a}{2^{M}})t]$ on Eq.(2) with $M=1,2,3,\cdots$, 
we can obtain 
\begin{equation}
H_{M}=(1-2^{-M})a^{+}a+g(a^{+}\sigma_{-}+a\sigma_{+})+g^{2}
\end{equation}
corresponding to the resonance conditions of $\Omega=2^{-M}$ under the RWA, and the eigenenergy in a 
Fock state representation is of the form  
\begin{equation}
E^{\pm}_{M}=(1-2^{-M})(n+\frac {1}{2})+\frac {\eta^{2}}{4}\pm\frac {1}{2}\sqrt{\eta^{2}(n+1)+(1-2^{-M})^{2}}
\end{equation}
with $n=0, 1, 2, \cdots$. Meanwhile,  performing a unitary
transformation of $\exp [-i(\frac {\Omega}{2K}\sigma_{z}+\frac {a^{+}a}{2})t]$ on Eq.(2) with $K=1,2,3, \cdots$, 
under the RWA resonance conditions of $\Omega=K$, we have 
\begin{equation}
H_{K}=\frac {K-1}{2K}\Omega\sigma_{z}+g(a^{+}\sigma_{-}+a\sigma_{+})+g^{2}
\end{equation}
whose eigenenergy in a Fock state representation is of the form
\begin{equation} 
E^{\pm}_{K}=\frac {\eta^{2}}{4}\pm\frac {1}{2}\sqrt{\eta^{2}(n+1)+(K-1)^{2}}
\end{equation}
with $n=0, 1, 2, \cdots$.

As shown in the last section, although our method can in principle present all series solutions, we merely present 
the specific forms of the non-RWA solution for two simplest cases. Here we will use those two cases for a comparison
with Eqs.(10) and (12). For the case (1), we find that the two solutions for $E=1\pm\epsilon$, along with two different 
restricted relations respectively, actually correspond to the same expression 
\begin{equation}
E=\frac {1}{2}+ \frac {1}{2}\eta^{2}+\frac {1}{8}\Omega^{2}. 
\end{equation}
However, for the case (2), the situation is somewhat complicated. The
direct algebra shown in Appendix presents us four solutions of E for this case, whereas the specific calculation 
shows that the solutions of Eqs.(A3) and (A4) are actually identical.

In Figs.2 and 3, the solutions under the RWA are compared with those for the non-RWA treatment in the case of 
$\Omega=$0.5 and 3.0. We can find that, in the case of small values of $\Omega$ and $\eta$, some of the solutions 
under the RWA can approach those without the RWA. It is physically reasonable because the RWA is only valid for the 
WER and LDL. However, although the solutions without the RWA are exact,  only a few of these solutions can be obtained 
for a certain termination of the series solution. We can not obtain all the particular solutions of the system unless 
we make the termination of the series solution for almost infinite times  up to the case with infinite series terms. 
In contrast, the RWA treatment can present general solutions, 
like Eqs.(10) and (12). Nevertheless, as it merely retains some resonance terms 
in the Hamiltonian, a specific solution under the RWA only corresponds to a specific value of $\Omega$. Moreover, 
the figures tell us that, only when the value of $\eta$ is not taken to be larger than 0.1, can we consider that the 
solutions under the RWA and without the  RWA are in good agreement in the case of the WER. The RWA description is 
obviously inaccurate for the case beyond the LDL although $\Omega$ is less than 1.0 in this case. 
When the value of $\Omega$ is larger,  the difference between the RWA case and 
non-RWA one becomes great, as shown in  Figs.3. It is difficult to find any correspondence for the two cases.   
However, there are some crossing points between the RWA  solutions and non-RWA ones. It means at  certain values of $\eta$, 
the solutions for eigenenergies in these
two cases can be the same. In physics, it can be considered that the non-RWA solutions are some isolated ones,  merely corresponding to 
those of up internal levels of the ion. One may not expect to obtain the general 
eigenenergies with our non-RWA treatment. However, the demonstration of above series solutions means the existence 
of the finite dimensional invariant subspace of some
operators. Particularly from the simple forms of the eigenenergies, one may
expect that the quantum KAM theorem should be applicable for
the SER problem$^{[20]}$.
 
Before ending our discussion of the usefulness of non-RWA solutions, it is also of interest to present a practical example
of preparing the nonclassical motional states of the ion with our solutions. As referred to in Sec.I, there 
is a considerable interest in investigating the SER, particularly for rapidly preparing nonclassical 
motional states of the ion. In fact, with the results in the present paper, 
one can also 
generate some particular nonclassical states through suitable adjustment of 
the Rabi frequencies and proper measurements. For example, consider the simplest case, i.e., the case (1) in Sec II.
For the case of the resonance ($\epsilon=0$), we have $E_{1}=E_{1}^{'}$ with the same
$\Omega$ and double degenerate eigenfunctions $\Psi (\alpha)$ and $\Psi^{'}(\alpha)$.
Transforming our results into the Schr\"odinger representation will yield the states $\Phi (t)=\exp{(-iE_{1}t+
i\omega_{l}\sigma_{z}t/2)}UV \Psi(\alpha)$ and $\Phi^{'} (t)=
\exp{(-iE_{1}t+i\omega_{l}\sigma_{z}t/2)}UV \Psi^{'}(\alpha)$, respectively. In the case of the LDL($g\ll 1$), 
 the Rabi frequency $\Omega$ would be $2\sqrt{1-4g^{2}}\sim 2$. So we have $c_{0}\sim \frac {g^{2}}{g^{2}-1}b_{0}$, 
 $b_{1}\sim \frac {g}{g^{2}-1}b_{0}$,
$c^{'}_{0}\sim \frac {g^{2}}{g^{2}-1}b^{'}_{0}$ and $b^{'}_{1}\sim \frac {g}{g^{2}-1}b^{'}_{0}$.
By controlling the evolution time $t=4\pi/\omega_{l}$, and measuring the excited state of
the ion, we can obtain a displaced even coherent state, a kind of 
Schr\"odinger-cat states
\begin{equation}
\Phi^{M}\sim \frac {1}{\sqrt{2}}D(\beta)(|i\frac {\eta}{2}>+|-i\frac {\eta}{2}>)=\frac {1}{\sqrt{2}}(|i\eta>+|0>)
\end{equation}
where the relation between the coherent state and Fock state has been used.
As the Rabi frequency approaches 2, such a preparation is obviously 
outside the WER.

\section{conclusion}

As $e^{g\alpha}\alpha^{n}$ corresponds to the displaced Fock states $|n,g>$, the eigenfunctions of the system
we obtained are infinite superposition of displaced Fock states. Therefore, it is understandable that we can
not obtain similar analytical results by general Fock state expansion. The obvious advantage of our approach 
is the possibility to obtain some relations between parameters of the system corresponding to certain 
nonclassical states by truncating the series expansion step by step. So as
long as we can reach these conditions
experimentally, the different nonclassical states predicted by our theory would be obtained.

In summary, the Raman interaction between a trapped ultracold ion and two
travelling wave lasers has been treated analytically 
in an interaction representation, without consideration of the RWA and the 
limitation for both the LDL and WER. Although the coherent state
expansion technique was used previously for treating a similar problem, and 
the solutions we presented here are some isolated ones under certain
conditions related to $\Omega$, $\eta$ and $\Delta$, the present investigation
is more general and exact so that we can study the ion trap problem in a wide 
range of parameters and compare our results with other approximate works or numerical 
solutions in this respect as we did in the present paper and for the cavity-atom problem$^{[21]}$.
 More importantly, 
the Raman process has been widely used for the laser-cooling in current ion trap 
experiments, and the work outside the WER is attracting much interest of 
experimentalists. One possible way$^{[10]}$ to experimentally realizing the SER is that the ion is 
first cooled within the LDL and 
under the WER, and then the trap frequency is decreased by opening the trap adiabatically 
so that  the ratio of the Rabi frequency to the trap frequency is increased to a large number.
Therefore, we believe that the difficulties in realizing the SER will soon be overcome, and 
 our work would be helpful for any possible future exploration of the
quantum properties of the ion-trap system outside the WER.

The author sincerely thanks M.A.Kornberg for his carefully reading the manuscript. 
The work is partly supported by National Natural Science Foundation of China.

\section{appendix}

For obtaining a more specific expression of the eigenenergy for the case (2) in Sec.II, we
can suppose $X=\epsilon - g^{2}$ for the case of $z=g$ and $Y=\epsilon +g^{2}$ for $z=-g$.
Direct algebra on Eq.(7) can present us following two second-order differential equations
$$AX^{2}+BX+C=0~~~~~~~~~~~~~~~~~~~~~~~~~~~~~~~~~~~~~~~~~~~~~~~~~~~~~~~~~~~~(A1)$$ and 
$$AY^{2}-BY+C=0~~~~~~~~~~~~~~~~~~~~~~~~~~~~~~~~~~~~~~~~~~~~~~~~~~~~~~~~~~~~(A2)$$ 
with $A=8(1-g^{2})$, $B=(3+\frac {\Omega}{2})(2+\Omega)(2-\frac {\Omega}{2})-3(\frac {\Omega}{2}-2
+\frac {\Omega^{2}}{4}+2g^{2})-28g^{2}-(3+\Omega)(2+\frac {\Omega}{2}-\frac {\Omega^{2}}{4}-2g^{2})$ and
$C=-[(\frac {\Omega}{2}-2+\frac {\Omega^{2}}{4}+2g^{2})(1+\frac {\Omega}{2})(2-\frac {\Omega}{2})
+20g^{2}-6g^{2}(\frac {\Omega^{2}}{4}+2g^{2})]$. So we can obtain the eigenenergies 
$$E^{\pm}_{1}=2+g^{2}+\frac {-B\pm\sqrt{B^{2}-4AC}}{2A}~~~~~~~~~~~~~~~~~~~~~~~~~~~~~~~~~~~~~~~~~(A3)$$ 
for the case of $z=g$ and 
$$E^{\pm}_{2}=2+g^{2}-\frac {B\pm\sqrt{B^{2}-4AC}}{2A}~~~~~~~~~~~~~~~~~~~~~~~~~~~~~~~~~~~~~~~~~(A4)$$
for the case of $z=-g$.

\newpage
\begin{center}{\bf Captions of the figures}\end{center}

Fig.1~~ Level scheme of the internal structure of the trapped ultracold
ion, where $|g>\leftrightarrow |e>$ is dipole forbidden.

Fig.2~~ Variation of E with respect to  $\eta$ in the case of 
$\Omega=0.5$, where solid curves from the bottom to  top are RWA solutions of Eq.(10), corresponding to 
$n=$0, 1, 2, 3, 4, 5 and 6, respectively, dot-dashed curve is the solution
of Eq(13) and dashed curves are from Eq.(A3) or Eq.(A4) (because the solutions of Eqs.(A3) and (A4) are
identical). (a) demonstrates $E^{+}$, and (b) is for $E^{-}$.

Fig.3~~ Variation of E with respect to  $\eta$ in the case of $\Omega=3.0$,  
where solid curves from the bottom to top are RWA solutions of Eq.(12) for the case of $E^{+}$, corresponding to 
$n=$0, 1, 2, 3, 4, 5 and 6, respectively, the dot-dashed curve is the solution
of Eq(13) and dashed curves are from Eq.(A3) for cases of $E^{\pm}$ respectively. 

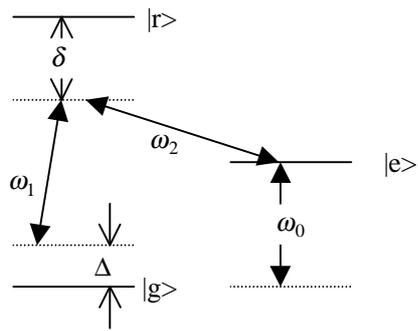
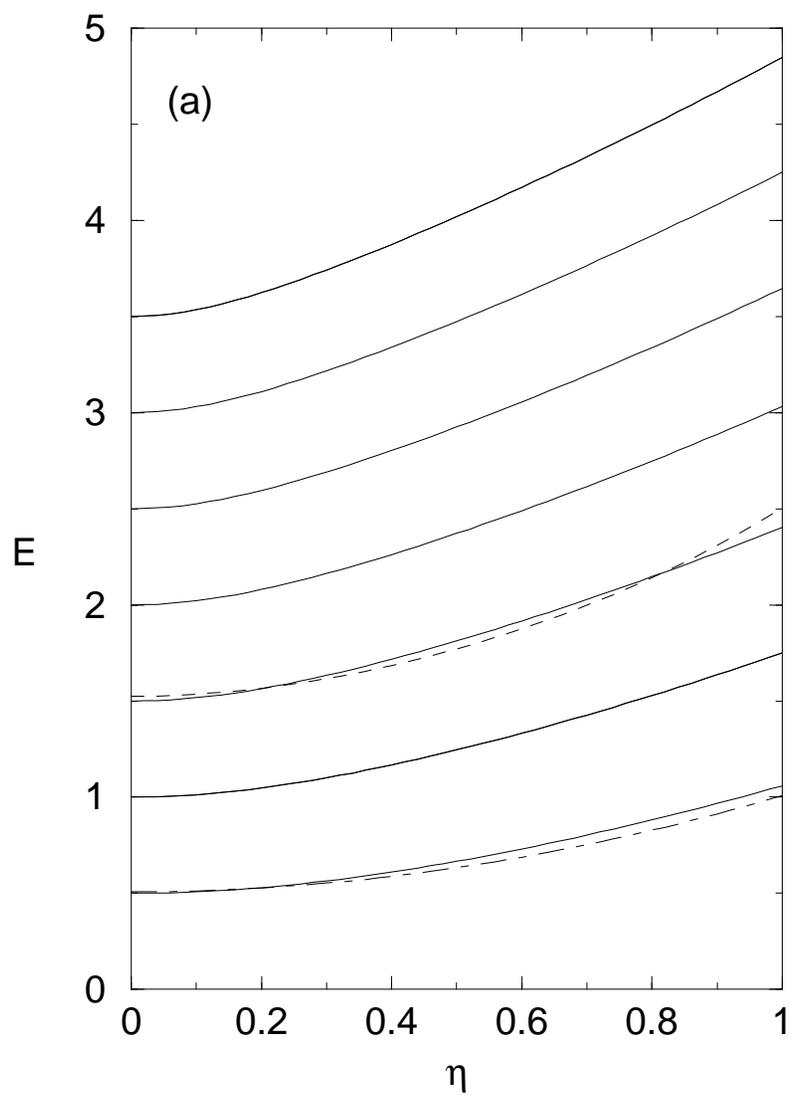
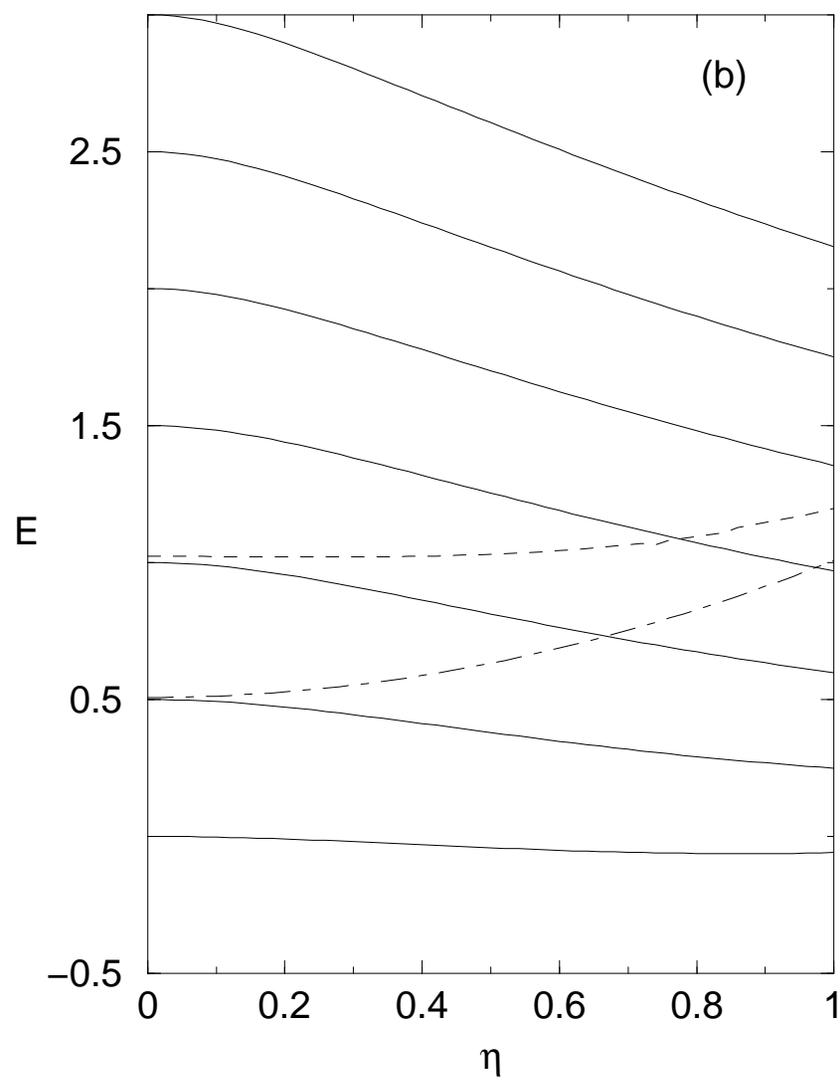
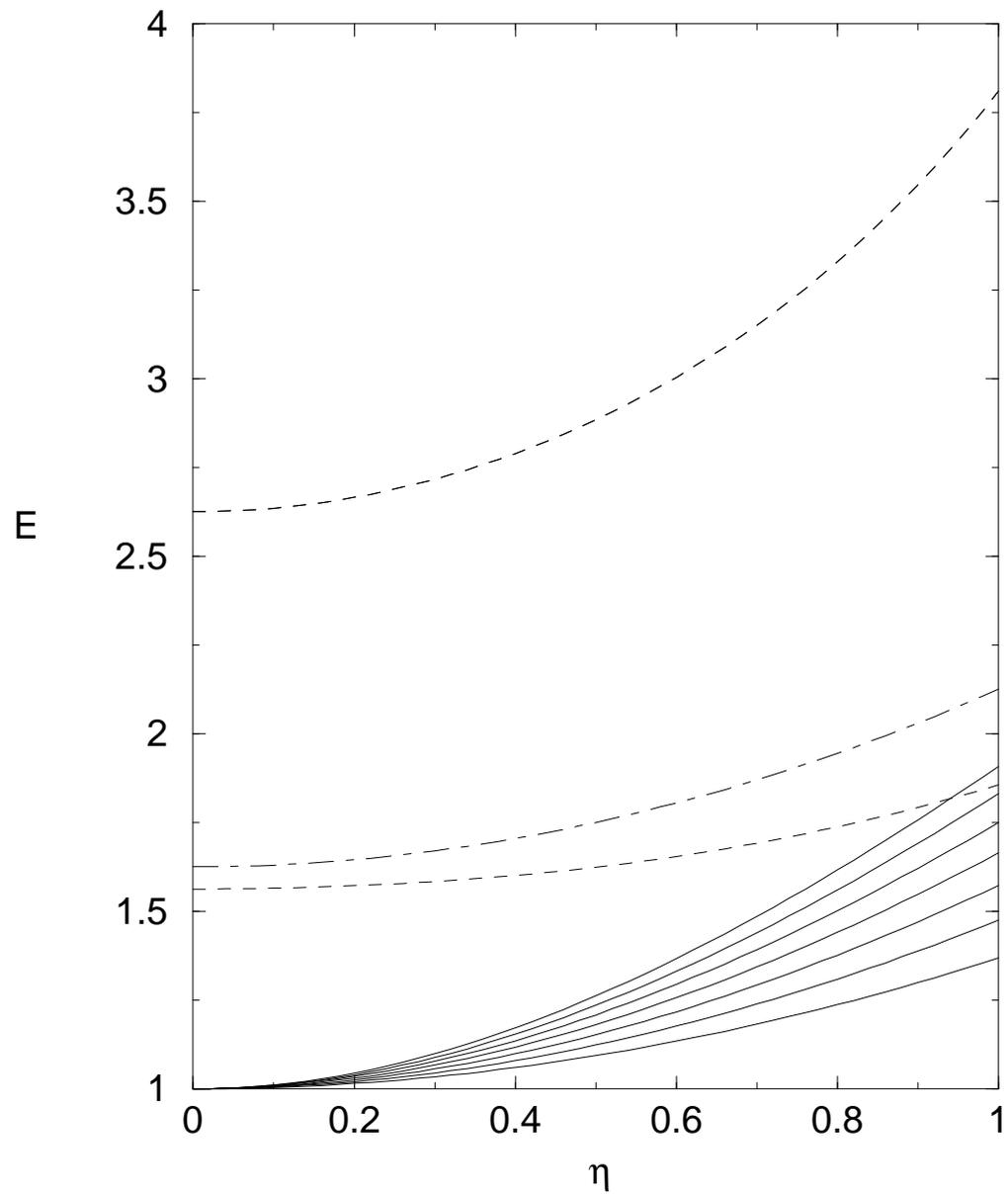
\end{document}